\documentclass[10pt,aps,prd,twocolumn,preprintnumbers,showpacs,nofootinbib,notitlepage,longbibliography,superscriptaddress]{revtex4-1}

\usepackage{amsmath, amssymb,braket}
\usepackage{color}
\usepackage{float}
\usepackage{graphicx}
\usepackage{subfigure}
\usepackage{natbib}
\usepackage{tikz}
\usepackage[colorlinks=true,citecolor=blue,urlcolor=blue]{hyperref}
\usepackage{xspace}

\begin{document}
\title{Resonant Pseudo-Dirac Dark Matter as a Sub-GeV Thermal Target}

\author{Nirmalya Brahma}
\affiliation{Department of Physics \& Trottier Space Institute, McGill University, Montr\'{e}al, QC H3A 2T8, Canada}

\author{Saniya Heeba}
\affiliation{Department of Physics \& Trottier Space Institute, McGill University, Montr\'{e}al, QC H3A 2T8, Canada}

\author{Katelin Schutz}
\affiliation{Department of Physics \& Trottier Space Institute, McGill University, Montr\'{e}al, QC H3A 2T8, Canada}

\begin{abstract}
	\noindent Dark matter (DM) could be a pseudo-Dirac thermal relic with a small mass splitting that is coupled off-diagonally to a kinetically mixed dark photon. This model, particularly in the sub-GeV mass range, is a key benchmark for accelerator searches and direct detection experiments. Typically, the presence of even a tiny fraction of pseudo-Dirac DM in the excited state around the time of recombination would be excluded by DM annihilation bounds from the cosmic microwave background (CMB); thus, viable thermal histories must typically feature an exponential suppression of the excited state. We revisit assumptions about the thermal history in the resonant regime, where the dark photon mass is slightly more than twice the DM mass (to within $\sim10\%$), leading to an $s$-channel resonance in the annihilation cross section. This resonance substantially reduces the couplings required for achieving the observed relic abundance, implying that in much of the parameter space, the DM kinetically decouples from the Standard Model well before the final DM relic abundance is achieved. We find that the excited state is not thermally depopulated in this regime. In spite of this, we find that the presence of the excited state does \emph{not} violate CMB bounds, even for arbitrarily small mass splittings. The present-day abundance of the excited state opens up the possibility of signatures that are usually not relevant for pseudo-Dirac DM, including indirect detection, direct detection, and self-interacting DM signatures.
\end{abstract}
\maketitle

\section{Introduction}
The origin of dark matter (DM) remains an elusive mystery. If the DM thermalizes with the Standard Model (SM) plasma in the early Universe, then thermal freeze-out provides a compelling explanation for the observed abundance of DM. In particular, thermal freeze-out is relatively insensitive to the initial conditions of the early Universe and the relevant couplings can be probed in a number of ways using direct detection, indirect detection, and collider observables~\cite{Jungman:1995df}. If the DM is lighter than the $\sim$GeV scale, then the Lee-Weinberg bound~\cite{lee1977cosmological} implies that the mediator for DM-SM interactions cannot be a SM force carrier, which opens up the possibility of a ``dark sector,'' with auxiliary forces and matter fields beyond just DM. A simple, technically natural example of a new mediator is a dark photon that kinetically mixes with the SM photon~\cite{Dienes:1996zr,Holdom:1985ag, Abel:2003ue,Abel:2004rp,Abel:2008ai,Acharya:2016fge,Acharya:2017kfi,Aldazabal:2000sa,Batell:2005wa,Gherghetta:2019coi}. Given the null detection of weak-scale DM thus far (see e.g. Refs.~\cite{Arcadi:2017kky,Gaskins:2016cha}), lighter dark sectors are of increasing interest to the community (see e.g. Refs.~\cite{Boehm:2003hm,Pospelov:2007mp,Feng:2008ya,Hochberg:2014kqa,Krnjaic:2015mbs}) and there are a range of new and proposed experimental methodologies that will be sensitive to these DM candidates~\cite{Essig:2011nj, Essig:2015cda,SuperCDMS:2018mne,DarkSide:2018ppu,DAMIC:2019dcn,EDELWEISS:2020fxc,Kahn:2021ttr}.

The standard thermal freeze-out mechanism for sub-GeV DM is subject to strong bounds from cosmic microwave background (CMB) anisotropies assuming annihilation through an $s$-wave process to visible SM particles. Even after freeze-out, DM can still annihilate at a sub-Hubble rate and inject considerable energy into the SM plasma near the time of recombination. Energy injection in the form of visible particles would observably modify the properties of the plasma even if the DM annihilations are extremely rare, since a $\sim$part-per-billion fraction of the DM annihilating would be enough energy injection to ionize all the atoms in the Universe. Considering the effects on CMB anisotropies as measured by \emph{Planck}, current bounds on DM annihilation rule out $s$-wave thermal freeze-out of DM below $\sim 10$~GeV, with the exact value depending on the SM final state~\cite{Slatyer:2015jla,Planck:2018vyg}.

A well-studied way to bring $s$-wave freeze-out into consistency with CMB constraints is by introducing a small mass splitting between non-degenerate DM states~\cite{Tucker-Smith:2001myb,Finkbeiner:2007kk, Arkani-Hamed:2008hhe, Cheung:2009qd, Chen:2009ab, Batell:2009vb,Graham:2010ca}. There is no symmetry that prevents Dirac fermions from splitting into two Majorana mass states, and this is easily realized in models where DM is charged under some new dark gauge symmetry at high energies which is broken at low energies \cite{Elor:2018xku,Duerr:2020muu}. The dark matter multiplet, consisting of $\chi_1$ and $\chi_2$, acquires a mass splitting $\delta = m_{\chi_2} - m_{\chi_1}$, which can be naturally small if the dark symmetry is approximate (for example, the small neutron-proton mass splitting is protected by an approximate isospin symmetry). In this work, we do not specify the origin of the mass splitting and treat it phenomenologically. However, we note that small mass splittings are generally easy to accommodate from a model-building perspective in situations with a small overall mass scale and small couplings \cite{Tucker-Smith:2001myb, Duerr:2019dmv, An:2011uq,Finkbeiner:2014sja, CarrilloGonzalez:2021lxm}. In this model, the couplings of DM with the dark photon are purely off-diagonal, i.e. the only vertex with the dark photon couples $\chi_1$ and $\chi_2$. For this reason, annihilation rates to SM final states can be significantly reduced because the leading-order tree-level annihilation process requires a large $\chi_2$ abundance, which may be thermally depleted like $e^{-\delta/T}$ at temperatures $T\lesssim \delta$ \cite{CarrilloGonzalez:2021lxm, Baryakhtar:2020rwy}. Given the $\sim$eV-scale temperatures that are relevant for recombination, mass splittings with $\delta \gtrsim 1$\,eV can be compatible with CMB constraints in parts of the parameter space (with larger mass splittings being unconstrained for a wider range of couplings and DM masses). 

Alternatively, if the DM annihilation occurs close to a pole in the cross-section (for instance when the mediator is close to twice the DM mass), then the relevant couplings to achieve the observed relic abundance can be lowered substantially~\cite{Griest:1990kh, Gondolo:1990dk}. If this pole is relevant for setting the DM abundance at early times but not during the recombination epoch, then the CMB bounds are relaxed because of the lower off-resonance annihilation rate to SM final states. The CMB bounds were carefully studied in Ref.~\cite{Bernreuther:2020koj} for the case of Dirac DM interacting with a dark photon with $m_{A'} \approx 2 m_\chi$. Meanwhile, the resonant regime for inelastic pseudo-Dirac DM has been studied primarily in the context of its signature at colliders given the modification to the predicted couplings~\cite{Izaguirre:2015yja,Feng:2017drg,Berlin:2018bsc,Duerr:2019dmv}. The cosmology of resonant pseudo-Dirac DM, on the other hand, has yet to be studied in detail. In particular, the substantial effects of early kinetic decoupling of the DM have been overlooked so far.

In this work, we perform a comprehensive study of the cosmology of pseudo-Dirac DM in the resonant regime. We find that even in the mildly resonant regime with $(m_{A^\prime} - 2 m_\chi)/m_{A^\prime} \sim 10\%$, pseudo-Dirac DM can have arbitrarily low mass splittings without violating limits from the CMB. Moreover, we find that in most of the parameter space the excited state has a high relic fraction. This provides a strong contrast to most pseudo-Dirac thermal histories which feature an exponential suppression of the excited state due to thermal depletion. Accordingly, the cosmology and astrophysics of this DM candidate are quite different from the usual pseudo-Dirac parameter space, as are the direct and indirect DM detection signatures. 

The rest of this article is organized as follows. In Section~\ref{sec:cosmo}, we review the model and the relevant processes that affect the cosmology of this DM candidate in the early Universe. In particular, we solve the Boltzmann equations for the density and temperature evolution of the DM states $\chi_1 $ and $\chi_2$. In Section~\ref{sec:cosmo_const}, we consider cosmological and astrophysical signatures including Big Bang Nucleosynthesis, the CMB, self-interacting DM (SIDM), and indirect detection. In Section~\ref{sec:experiment} we discuss prospects for detecting this DM candidate using terrestrial experimental methods. Discussion and concluding remarks follow in Section~\ref{sec:conclusion}

\section{Early Universe Behaviour}
\label{sec:cosmo}
\subsection{Pseudo-Dirac DM Parameter space}
We consider a light ($m_\chi \lesssim 10$~GeV) pseudo-Dirac DM model with its relic abundance set by annihilation to SM final states via a dark photon mediator. We focus on this mass range primarily because Dirac DM with $m_\chi \lesssim 10$~GeV is excluded by the CMB for $s$-wave freeze-out to visible final states~\cite{Planck:2018vyg}. In this model, the interaction terms are
\begin{align}
\mathcal{L} \supset \frac{\kappa}{2}F^\prime_{\mu\nu}F^{\mu\nu} + i g_\chi A^\prime_\mu \chi_2\gamma^\mu\chi_1\,,
\end{align}
where $\chi_{2,1}$ are the excited and ground states, respectively, that couple with interaction strength $g_\chi$ to a vector mediator $A^\prime$ that kinetically mixes with the SM photon with mixing parameter $\kappa$. We focus on parameter space where the mass splitting is much smaller than the DM mass, $\delta = m_{\chi_2} - m_{\chi_1} \ll  m_{\chi_1} $. In the following discussion, we denote the average mass of the two states as $m_\chi$. We are furthermore interested in the resonant regime where $m_{A^\prime} \approx m_{\chi_2} + m_{\chi_1} = 2m_{\chi_1} + \delta= 2 m_\chi$. We parameterize the proximity to resonance with the parameter 
\begin{equation}
    \epsilon_R = \frac{m_{A^\prime}^2 - s_0}{s_0} 
\end{equation}
where $s_0 = (m_{\chi_1}+m_{\chi_2})^2$. In this work, we consider $\epsilon_R \in [0.001,0.1]$. The lower limit is motivated by photodisassociation bounds coming from BBN as discussed in Sec. \ref{sec:bbn}. On the other hand, as $\epsilon_R>0.1$, we approach the non-resonant limit. We remain agnostic about the mechanism responsible for generating the mass splitting as well as the resonance, however the parameters we consider are self-consistent even in a minimal UV setup. For instance, we could consider a complex dark Higgs with a dark charge of 2 interacting with a Dirac fermion with a mass $m_D$~\cite{Elor:2018xku, Duerr:2020muu,Heeba:2023bik}. The breaking of the dark $U(1)$ symmetry through the vev of the dark Higgs, $v_D$, then results in both the mass term for the dark photon as well as a Majorana mass term for the Dirac fermion which generates the mass splitting. If the dark Higgs is heavy with $v_D \gg m_D$ then it will not participate in the dynamics that determine the DM thermal history. With this hierarchy in mind, having the dark photon mass near its resonant value pushes $g_\chi \sim m_D/v_D\ll1$, which we show below is consistent with setting the observed DM relic abundance in the resonant regime. Finally, the mass splitting is determined by the Yukawa coupling of the Dirac fermion $y_\chi$, with $\delta \sim y_\chi v_D$.

Though the dark photon is the mediator of this model, it can be resonantly produced on shell via inverse decays. The dark photon can subsequently decay invisibly,
\begin{equation}
   \Gamma_\text{DM} = \frac{g_\chi^2 m_{A'}}{12 \pi} \sqrt{1 - \frac{s_0}{m_{A'}^2}}  \left( 1 + \frac{s_0}{2m_{A'}^2} \right)\left(1 - \frac{\delta^2}{m_{A^\prime}^2}\right)^{3/2}\,,
\end{equation}
or visibly to SM final states, 
\begin{equation}
    \Gamma_\text{SM} = R(m_{A'})\Gamma_{\mu^+ \mu^-} + \sum_\ell \Gamma_{\ell^+ \ell^-} 
\end{equation}
where $R(m_{A'})$ is the empirically determined branching ratio of $\sigma(e^+ e^- \rightarrow \text{hadrons})/\sigma(e^+e^- \rightarrow \mu^+\mu^-) $ \cite{Ezhela:2003pp,ParticleDataGroup:2018ovx} at centre-of-mass energy $\sqrt{s} = m_{A'}$ and 
\begin{equation}
    \Gamma_{\ell^+ \ell^-} = \frac{\kappa^2 e^2 m_{A'}}{12 \pi} \sqrt{1 - \frac{4 m_\ell^2}{m_{A'}^2}}  \left( 1 + \frac{2 m_\ell}{m_{A'}} \right) .
\end{equation}
The total on-shell decay width is $\Gamma_{A'} \equiv \Gamma_\text{DM} + \Gamma_\text{SM} \equiv \Gamma_\text{DM} + \Gamma_{e^+ e^-}/B_e $, where $B_e$ is the branching ratio of the dark photon to electrons. 

\subsection{Relic Density Assuming Thermal Equilibrium}
\begin{figure}[t!]
\includegraphics[width=0.49\textwidth]{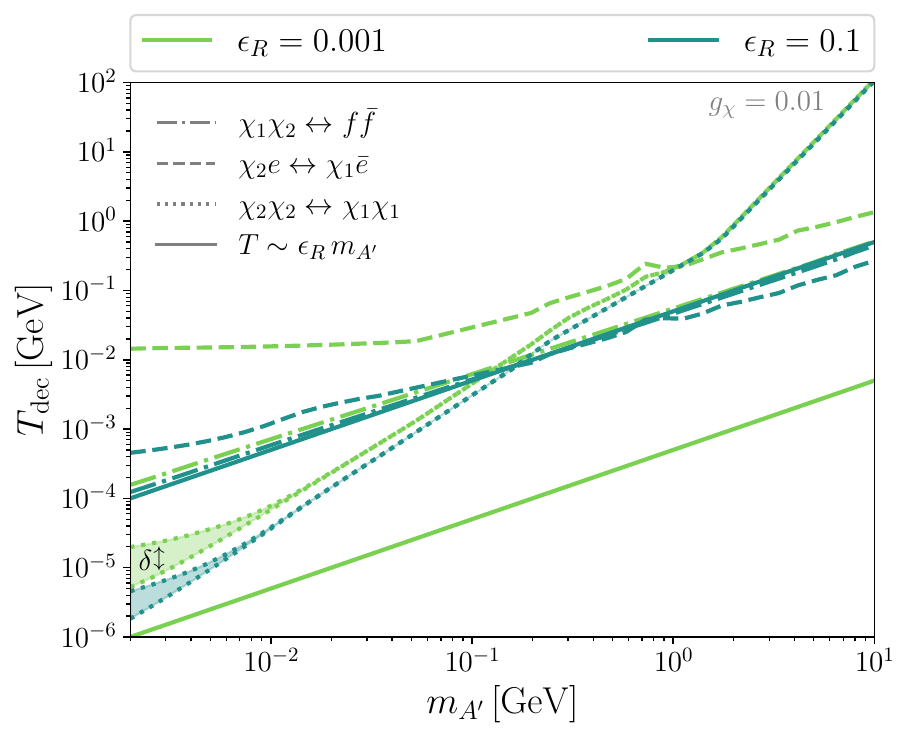}
\vspace{-0.3cm}        
\caption{The SM temperature as a function of the dark photon mass at which DM-SM scattering (dashed), DM-DM scattering (dotted) and DM-DM annihilation (dot-dashed) decouple for $\epsilon_R=0.001$ (light green) and $\epsilon_R=0.1$ (dark green), with $g_\chi=0.01$. The solid lines show when annihilation becomes resonant, with $T\sim \epsilon_R m_{A^\prime}$. For each parameter point in this plot $\kappa$ is chosen so as to obtain the observed DM relic abundance ignoring the effects of early kinetic decoupling. The shaded areas between dotted lines correspond to varying $\delta$ between $1-100\,\mathrm{eV}$, from bottom to top. In some parts of the parameter space, resonant depletion of the DM abundance occurs much later than other decoupling processes, indicating that the early decoupling can influence the subsequent relic DM abundance. }
\label{decouplingtemp}
\vspace{-0.3cm}     
\end{figure}
In our parameter region of interest, DM obtains its relic abundance when $s$-channel processes like $\chi_1\chi_2\to \text{SM SM}$ become inactive, usually well after chemical freeze-out for $\epsilon_R \ll1$. Using the formalism developed in Ref.~\cite{Feng:2017drg}, the corresponding thermally averaged cross-section can be written as 
\begin{align}
\label{eq:sigmavthav}
    \langle \sigma v \rangle  = \frac{2x}{K_2^2(x)}\int_0^{\infty}\mathrm{d\epsilon}\,\sigma v \sqrt{\epsilon}\,(1+2\epsilon)K_1(2x\sqrt{1+\epsilon})\,,
\end{align}
where $x = m_{\chi_1}/T$, $\epsilon = (s-s_0)/s_0$ is a dimensionless measure of the kinetic energy and 
\begin{align}
\label{eq:sigmav}
    \sigma v = F(\epsilon) \frac{m_{A^\prime}\Gamma_{A^\prime}}{(s-m_{A^\prime}^2)^2 + m_{A^\prime}^2\Gamma_{A^\prime}^2}
\end{align}
with,
\begin{align}
F(\epsilon) = &\frac{4\pi \kappa^2\alpha\alpha_D}{3s_0 m_{A^\prime}\Gamma_{A^\prime}}\frac{(3+2\epsilon)[(1+\epsilon)s_0 + 2m_e^2]}{(1+\epsilon)(1+2\epsilon)B_e(\sqrt{s_0(1+\epsilon)})}\nonumber\\
&\times[s_0(1+\epsilon)-4m_e^2]^{1/2}[s_0(1+\epsilon)-\delta^2]^{1/2}\,.
\end{align}
The thermally averaged cross-section can be further reduced to semi-analytic forms in the non-relativistic ($\epsilon \ll1$) and resonant ($\epsilon \approx \epsilon_R$) limits \cite{Bernreuther:2020koj}.

In order to calculate the \textit{total} DM relic density, one needs to solve the Boltzmann Equation for $\chi_1\chi_2\to \text{SM SM}$ , 
\begin{align}
\frac{\mathrm{d}Y_{\mathrm{tot}}}{\mathrm{d}x} = \frac{s}{H x}\langle\sigma v\rangle_\mathrm{eff} \left(Y_{\mathrm{tot}}^2 -Y_{\mathrm{tot},\, \mathrm{eq}}^2\right)\,,
\end{align}
where $Y_{\mathrm{tot}}=Y_{\chi_1}+Y_{\chi_2}$ is the total comoving density for DM with $Y_{\chi_1,\chi_2} = n_{\chi_1,\chi_2}/s$, $s$ is the entropy density, and the effective thermally averaged cross-section is \cite{Griest:1990kh, Feng:2017drg}
 \begin{align}
    \langle \sigma v \rangle_\mathrm{eff} = \frac{2(1+\delta/m_\chi)^{3/2}e^{-x\delta/m_{\chi_1}}}{(1+(1+\delta/m_{\chi_1})^{3/2}e^{-x\delta/m_{\chi_1}})^2}\langle \sigma v\rangle .
\end{align}
The total DM relic density is then given by
\begin{align}
    \label{eq:Oh2}
    \Omega_{\mathrm{DM}}h^2 = 8.77 \times 10^{-11} \left[  \int_{x_f}^{\infty}\mathrm{d}x\frac{\langle\sigma v\rangle_\mathrm{eff}}{x^2} g_\ast^{1/2}\right]^{-1}\,,
\end{align}
where $g_\ast$ correspond to the effective relativistic degrees of freedom in the early universe. It can be seen from Eqs.~\eqref{eq:sigmavthav} and \eqref{eq:sigmav} that the cross-section is resonantly enhanced when $\chi_1$ and $\chi_2$ have enough energy to produce the $A^\prime$ on shell, leading to very efficient annihilation at a temperature, $T\sim \epsilon_R m_{\chi_1}$.
Due to the resonant enhancement, as $\epsilon_R \to 0$ even very tiny couplings are able to efficiently deplete the DM in the early Universe to obtain the observed DM relic density.

\subsection{Early Kinetic Decoupling}
A crucial caveat to the relic density calculation detailed above is that it assumes that the DM and SM remain in kinetic equilibrium while annihilations are active (including after chemical freeze-out) through scattering processes, primarily off of electrons $\chi_1 e\leftrightarrow\chi_2 {e}$. Since this is a $t$-channel process, it does not benefit from the same resonant enhancement as the $s$-channel annihilations. Therefore, when the couplings between the two sectors are small, the assumption of kinetic equilibrium may no longer hold and DM can kinetically decouple well before DM annihilation $\chi_1\chi_2\to \mathrm{SM}\,\mathrm{SM}$ hits its resonance (compare e.g. the dashed and solid lines in Fig.~\ref{decouplingtemp}). As a result, to accurately calculate the DM relic density, one needs to solve a coupled system of differential equations tracking the evolution of both the DM number density as well as the dark sector temperature, $y (x) \equiv m_{\chi_1} T_\mathrm{DM} s^{-2/3}$~\cite{vandenAarssen:2012ag,Binder:2017rgn, Binder:2021bmg},
\begin{figure*}[t!]
\centering
\includegraphics[width=0.7\textwidth]{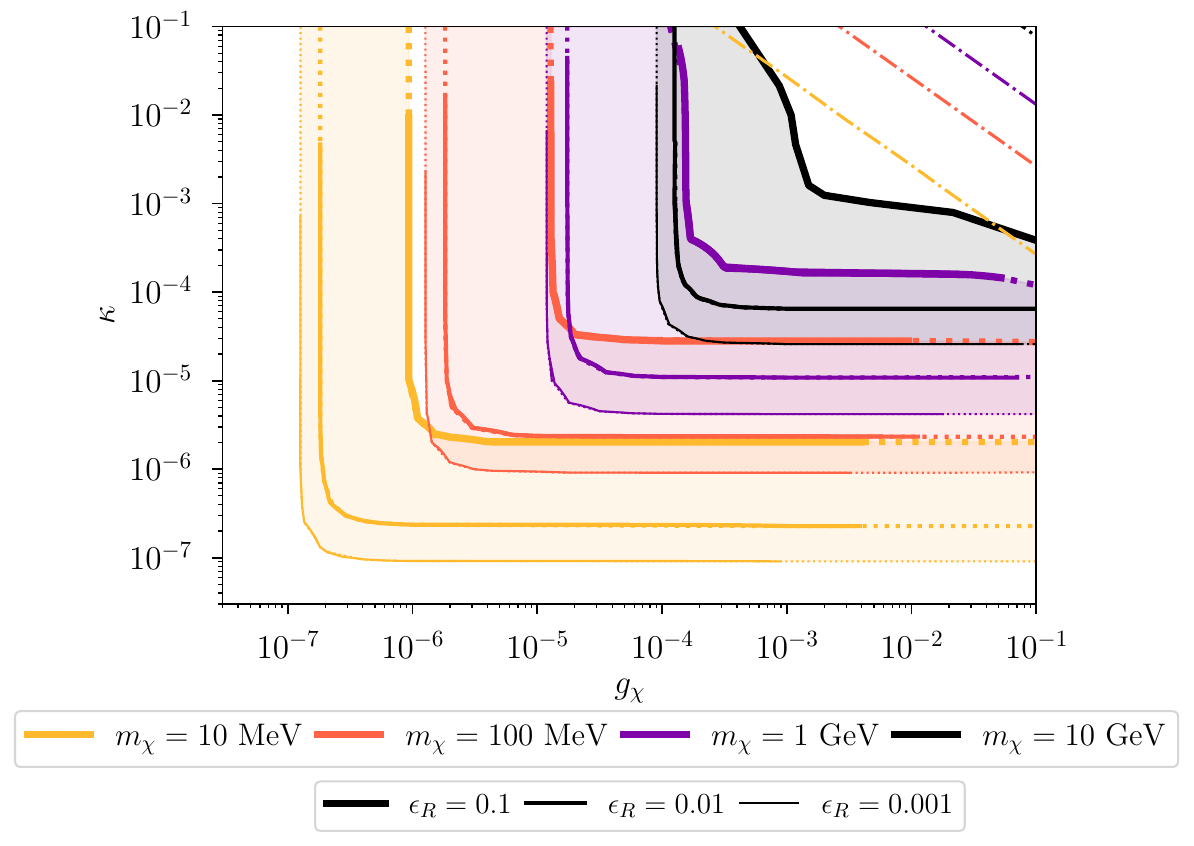}
\vspace{-0.2cm}
\caption{The couplings that yield a DM abundance that matches the observed relic density for various DM masses and values of $\epsilon_R$. Dashed lines indicate CMB annihilation constraints on those couplings, while solid lines are consistent with the CMB. In this parameter space, the smaller of the two couplings (corresponding to whichever decay channel is the bottleneck) determines the relic abundance, with the larger coupling being irrelevant. Small deviations from this behaviour occur at large values of $m_\chi$ and $\epsilon_R$ where early kinetic decoupling has a significant effect on the relic abundance. These trends contrast with the parameter space for standard thermal freeze-out where the product of couplings is the most relevant for setting the relic abundance. For comparison, we show the couplings for thermal freeze-out for $m_{A^\prime} = 3 m_\chi$ (off resonance) as dot-dashed lines.
}
\vspace{-0.2cm}
\label{kappavsgx}
\end{figure*}
\begin{align}
\label{eq:BE}
    \frac{Y^\prime}{Y} &= \frac{sY}{x\tilde{H}}\left(\frac{Y^2_\mathrm{eq}}{Y^2}\langle\sigma v\rangle - \langle \sigma v \rangle_{\mathrm{neq}}\right)\,\\
    \frac{y^\prime}{y} &= \frac{\gamma(T)}{x\tilde{H}}\left(\frac{y_\mathrm{eq}}{y}-1\right) + \frac{sY}{x\tilde{H}}\left(\langle\sigma v\rangle_{\mathrm{neq}}-\langle\sigma v\rangle_{2,\,\mathrm{neq}}\right) \label{eq:yDM}\nonumber \\
    +& \frac{s Y}{x \tilde{H}}\frac{Y^2_\mathrm{eq}}{Y^2}\left[\frac{y_\mathrm{eq}}{y}\langle\sigma v\rangle_2 - \langle\sigma v\rangle\right] + \frac{H}{x\tilde{H}}\frac{\langle p^4/E^3\rangle_\mathrm{neq}}{3T_\mathrm{DM}}
\end{align}
where $\tilde{H}$ is  the normalised Hubble rate as defined in Ref.~\cite{Binder:2017rgn} and the subscript ``neq'' denotes that the corresponding thermal average is over the DM phase distribution at a temperature $T_{\mathrm{DM}}$ distinct from the SM temperature $T$, i.e., assuming DM is not necessarily in kinetic equilibrium with the SM. Additionally, $\langle\sigma v \rangle_2$ is a temperature-weighted analog of the usual thermally averaged annihilation cross-section $\langle \sigma v \rangle$, as defined in Ref.~\cite{Binder:2017rgn}, and $\gamma(T)$ is the DM-SM momentum transfer rate which is a measure of DM-SM elastic scattering. Terms involving scattering and annihilation can both keep the DM temperature coupled to the SM. These Boltzmann equations have been extensively studied for the case of elastically decoupling relics ~\cite{vandenAarssen:2012ag,Binder:2017rgn,Binder:2018msg,Binder:2021bmg}. For inelastic DM models, the corresponding Boltzmann equations may have an additional dependence on the mass-splitting $\delta$, which would appear in various cross sections, and there could also in principle be separate thermal evolution of the ground and excited state species. However, in our parameter region of interest, $\delta \ll m_{\chi}$ and $\delta$ is also much smaller than the decoupling temperatures for all relevant processes. Therefore, we explicitly find that the thermal history of this model behaves, to a very good approximation, as a strictly Dirac model during the temperatures relevant for freeze-out. Therefore, to calculate the relic pseudo-Dirac DM density, we use the publicly available Boltzmann solver \texttt{DRAKE} ~\cite{Binder:2021bmg} modified for a resonant Dirac DM model. 

\subsection{Parameter Space}
The couplings required to reproduce the observed DM abundance are shown in Fig. \ref{kappavsgx} for different values of $m_{\chi}$ and $\epsilon_R$. Note that $\delta$ is generally much smaller than all the temperatures that are relevant for setting the relic abundance, and therefore it is irrelevant in determining the couplings. As expected, for a given DM mass, smaller values of $\epsilon_R$ (indicated by thinner lines) result in a larger resonant enhancement in the annihilation cross-section, and corresponds to smaller couplings reproducing the relic density, thereby shifting the lines downwards and to the left. For a fixed value of $m_{\chi}$ and $\epsilon_R$, the shape of the curve in the $g_\chi-\kappa$ plane can be explained by considering the thermally averaged cross-section in the limit $\epsilon_R \ll 1$. In this case, as was pointed out in Ref.~\cite{Bernreuther:2020koj}, the slower decay ($\Gamma_\text{SM}$ vs. $\Gamma_\text{DM}$) is the bottleneck in terms of determining the final DM abundance,
\begin{align}
    \Omega_\mathrm{DM}h^2 \propto \frac{\Gamma_{A^\prime}}{\kappa^2g_\chi^2}\,.
    \label{eq:oh2scale}
\end{align}
This implies that for $g_\chi \ll \kappa$ ($\kappa \ll g_\chi$), the relic density becomes independent of $\kappa$ ($g_\chi$) resulting in the asymptotic behavior seen in Fig. \ref{kappavsgx}. 

In the limit $\epsilon_R \rightarrow 0$, we find that the relic density obtained using the coupled system of Boltzmann Eqs.~\eqref{eq:BE}-\eqref{eq:yDM} differs from the standard Boltzmann treatment (i.e. assuming identical SM and DM temperatures) by at most a factor of $\sim 2$. For $\epsilon_R\sim1$, one would naively expect the difference to be even smaller since we are not only further off-resonance but are also pushed toward larger couplings where the expectation is that kinetic equilibrium should be maintained more easily. However, we find that {the difference between the two treatments can be as large as an order of magnitude in the relic density} in the region $\epsilon_R \sim 0.1$. This can be attributed to the deviation of the DM temperature from the SM temperature and was earlier discussed in the context of scalar singlet DM in Ref. \cite{PhysRevD.96.115010}. In particular, for $\epsilon_R \ll 0.1$, the deviation is very small and positive, $T_\mathrm{DM} \gtrsim T$ whereas for $\epsilon_R \sim 0.1$ the deviation is large and negative, $T_\mathrm{DM} \ll T$. Although one can numerically estimate the sign and magnitude of this deviation by studying the interplay of the various terms on the right hand side of Eq.~\eqref{eq:yDM}, they can also be understood qualitatively by considering the underlying DM phase space during and after chemical freeze-out. 

\begin{figure}[t!]
    \centering    \includegraphics[width=0.49\textwidth]{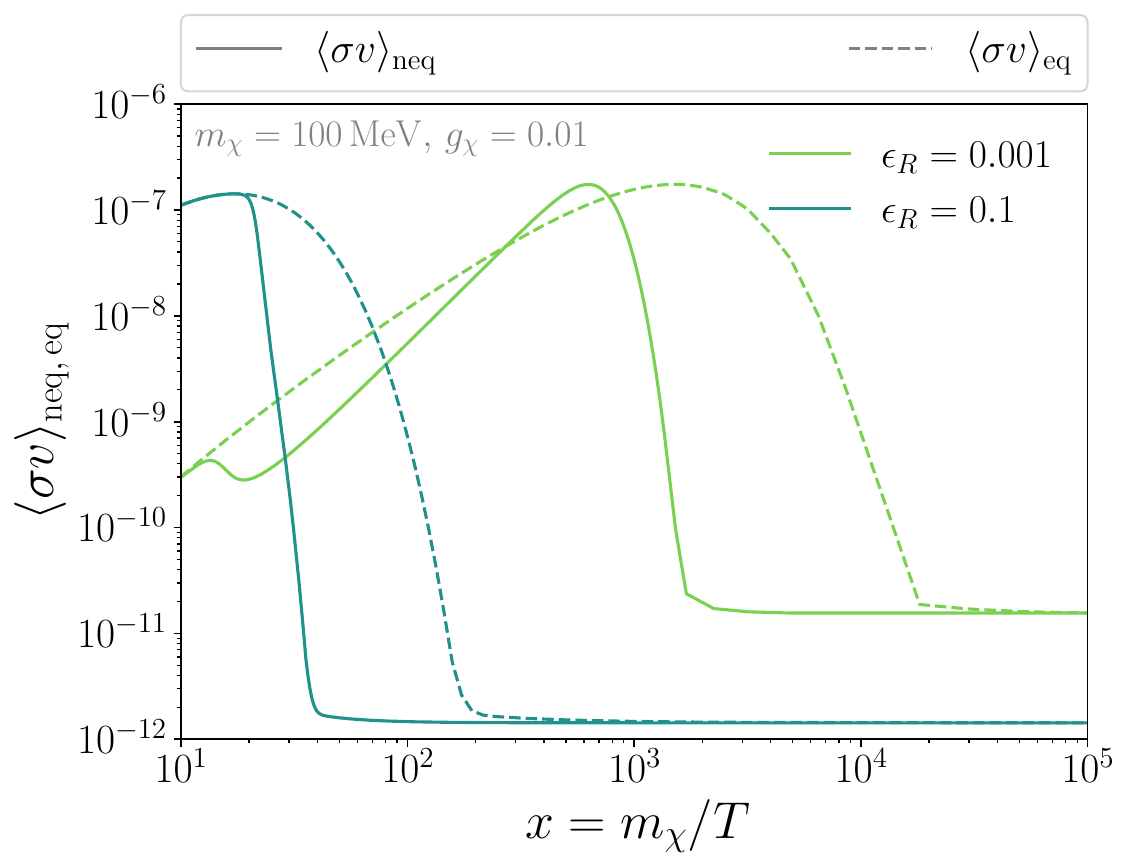}
    \vspace{-0.4cm}
    \caption{The thermally averaged annihilation cross-section for kinetically decoupled (solid) and kinetically coupled (dashed) DM as a function of $x$ (defined with respect to the SM temperature, $T$) for $\epsilon_R=0.1$ (dark green) and $\epsilon_R=0.001$ (light green) and $m_\chi=100\,\mathrm{MeV}$. We fix $g_\chi=0.01$ and choose $\kappa$ such that we reproduce the observed DM abundance after solving Eqs.~\eqref{eq:BE}-\eqref{eq:yDM}. Early kinetic decoupling can suppress or enhance resonant annihilation at given temperature.}
    \vspace{-0.2cm}
    \label{fig:epsRpt1} 
\end{figure}

Under the assumption that DM has already kinetically decoupled, the final DM density can be assumed to be proportional to the annihilation cross-section averaged over the DM temperature, $\langle\sigma v\rangle_\mathrm{neq}$ (in analogy with Eq.~\eqref{eq:Oh2}) 
\begin{align}
\label{eq:omegahneq}
    \Omega_\mathrm{DM}h^2 \propto \left[\int_{x_f}^\infty \mathrm{d}x \frac{\langle\sigma v\rangle_\mathrm{neq}}{x^2}\right]^{-1}\,.
\end{align} 
In general, as $\epsilon_R\rightarrow 0$, the DM particles need only very little momentum to hit the resonance and annihilate efficiently, implying that resonant annihilation depletes the low-momentum tail of the DM distribution and shifts the average DM momentum (and therefore the temperature) to larger values. During chemical freeze-out, since the average DM momentum is already large, only a small fraction of DM particles can annihilate resonantly. The increase in the DM temperature further decreases the available phase space for resonant annihilation and therefore decreases $\langle\sigma v\rangle_\mathrm{neq}$. This results in the small dip in $\langle\sigma v\rangle_\mathrm{neq}$ for $\epsilon_R =0.001$ around chemical freeze-out $x_f \sim 20$ seen in Fig.~\ref{fig:epsRpt1} and corresponds to reducing the efficiency of DM annihilations and increasing its abundance. After DM has chemically decoupled, its temperature now redshifts as matter and therefore DM cools much faster than the SM, increasing the relative number of DM particles with low momentum. As a result, resonant annihilation which is active at $T_\mathrm{DM}\sim \epsilon_R m_{\chi}$ happens at slightly earlier times (since $T_\mathrm{DM}\ll T$) compared to the kinetically coupled case when they occur at $T \sim \epsilon_R m_{\chi}$. Hence, resonant annihilation is more efficient in depleting the DM (due to the $x$ dependence of Eq.~\eqref{eq:omegahneq}). Since these two effects change the relic density in opposing ways, the final relic density is only slightly different from the kinetically coupled case. This is especially true given that the $1/x^2$ weighting in the integral of Eq.~\eqref{eq:omegahneq} ensures that the most substantial contributions to the relic abundance come from early times before the difference between $\langle\sigma v\rangle_\mathrm{eq}$ and $\langle\sigma v\rangle_\mathrm{neq}$ becomes too large. 

For $\epsilon_R\sim 0.1$, on the other hand, DM particles need larger momentum to annihilate resonantly and therefore resonant annihilations shift the average DM momentum (and temperature) to smaller values. Additionally,  resonant annihilation is active exactly during chemical freeze-out, $x_\mathrm{DM}\sim\epsilon_R^{-1}\sim x_f$. This implies that if DM is kinetically decoupled, the depletion in the large-momentum DM phase space effectively turns off resonant annihilations quite quickly as shown by the dark green line in Fig.~\ref{fig:epsRpt1}. Furthermore, the $x^2$ scaling of Eq.~\eqref{eq:omegahneq} in this regime enhances the difference in the total relic abundance since $\langle\sigma v\rangle_\mathrm{eq}$ and $\langle\sigma v\rangle_\mathrm{neq}$ differ substantially around the time of chemical freeze-out. As a result, the relic density is significantly altered: DM is over-produced and larger couplings are required to obtain the observed DM abundance. 

\subsection{Late-time Abundance of the Excited State}
The relative fraction of excited-state particles, $f^\ast = n_{\chi_2}/n_{\chi_1}$ is a key quantity in determining the impact of late-time DM behaviour in cosmological environments as well as terrestrial experiments, as discussed further in Sections~\ref{sec:cosmo_const} and~\ref{sec:experiment}. Even if DM is symmetrically produced in the ground and excited states, the ground and excited states can inter-convert through processes within the dark sector or through scattering processes with the SM as long as their rates exceed the Hubble expansion rate. In particular, as long as chemical equilibrium is maintained \textit{within} the dark sector, the excited state number density is given by $n_{\chi_2} \sim n_{\chi_1}e^{-\delta/T_\mathrm{DM}}$. Chemical equilibrium in the dark sector can be maintained through DM-SM scattering, which also maintains kinetic equilibrium with the SM, $\chi_1 e\leftrightarrow \chi_2 {e}$, or through DM up/down-scattering, $\chi_1\chi_  1\leftrightarrow\chi_2\chi_2$. The fractional abundance of a cosmologically stable excited state at late times is determined the DM temperature when it chemically decouples, $f^\ast = n_{\chi_2}/n_{\chi_1} \approx e^{-\delta/T_\mathrm{chem}}$, where $T_\mathrm{chem}$ is determined by whichever of the processes listed above decouples last, $T_\mathrm{chem}= \min[T_{\chi e},\,T_{\chi\chi}]$,
where
\begin{align}
    \left.\frac{n_e\langle\sigma v\rangle_{\chi_2 e\to\chi_1 \bar{e}}}{H}\right\vert_{T=T_{\chi e}} \sim 1\,,\\
     \label{eq:xxdec}\left.\frac{n_{\chi_2}\langle\sigma v\rangle_{\chi_2 \chi_2\to\chi_1 \chi_1}}{H}\right\vert_{T=T_{\chi \chi}} \sim 1.
\end{align}
Here, $n_e$ is the electron number density, $n_{\chi_2} = n_{\chi_1}e^{-\delta/T_\mathrm{DM}}$ is obtained by scaling back the present-day DM abundance $n_{\chi_1} \sim T_\mathrm{eq}T^3/m_{\chi_1}$, and the relevant cross-sections are from Ref.~\cite{CarrilloGonzalez:2021lxm}. If $\chi_2\chi_2\leftrightarrow\chi_1\chi_1$ decouples after $\chi_2 e\leftrightarrow \chi_1 {e}$, two temperature scales enter in Eq.~\eqref{eq:xxdec}, the SM temperature $T$ that largely determines the Hubble rate, and the DM temperature which is a function of the temperature at which DM decouples from the SM, $T_\mathrm{DM} \sim T^2/T_{\chi e}$.

\begin{figure}[t!]
\includegraphics[width=0.49\textwidth]{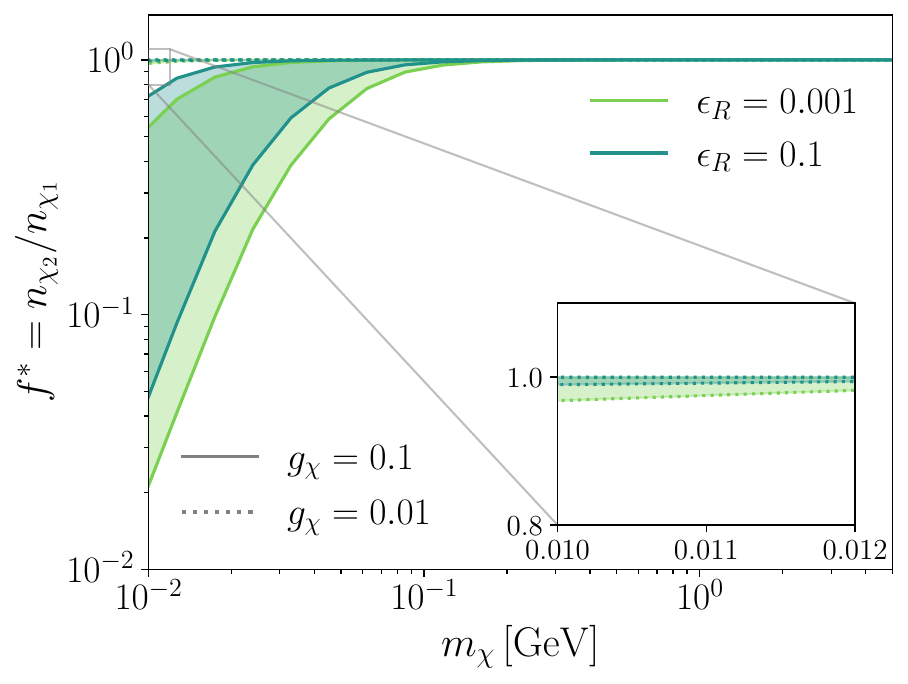}
\vspace{-0.3cm}        
\caption{The relative number density of the excited and ground states, $f^* = n_{\chi_2}/n_{\chi_1}$, as a function of the DM mass. The different lines correspond to $g_\chi=0.1$ (solid) and $g_\chi=0.01$ (dotted). The two colours correspond to two different values of $\epsilon_R$ and the shaded areas correspond to varying $\delta$ between $1-100~\rm{eV}$ from top to bottom. Due to the early decoupling of the processes that would deplete the excited state, the excited state remains abundant at late times in most of the parameter space, in contrast to most other pseudo-Dirac DM thermal histories.} 
\label{fig:fstar}
\vspace{-0.3cm}     
\end{figure}
In the standard thermal history for inelastic DM, one finds that $T_\mathrm{chem} \lesssim \delta$
owing to the large DM-DM and/or DM-SM couplings which ensures chemical equilibrium in the dark sector is maintained until late times. This results in a strong suppression of the excited state at late times, $f^\ast \sim \mathcal{O}(10^{-4})$ \cite{Baryakhtar:2020rwy,Fitzpatrick:2021cij}. However, the small couplings present in our parameter space result in $T_\mathrm{chem} \gg \delta$, and therefore a similar abundance of the ground and excited state, $f^\ast \sim 1$ in much of the parameter space. This is represented in Fig. \ref{fig:fstar}, in which we show the relative abundance of the ground and excited states as a function of the DM mass for different values of $\epsilon_R$ and $g_\chi$. Note that $f^\ast \sim 1$ for $g_\chi \ll 0.1$ for all DM masses of interest. For $g_\chi \sim 0.1$, we find a suppression of the excited state to $f^\ast \sim 0.01$ when $m_{\chi} \lesssim 100 \,\mathrm{MeV}$.

\section{Cosmological and Astrophysical Constraints}
\label{sec:cosmo_const}
Due to the high late-time abundance of the excited state, this thermal history has unique signatures compared to other pseudo-Dirac DM thermal histories. In this Section, we determine the qualitatively new behavior in astrophysical systems caused by the presence of the excited state and estimate the resulting constraints on the model as inferred from existing measurements. 
\subsection{BBN}
\label{sec:bbn}
Sub-GeV DM may significantly impact the abundance of light elements in the Universe produced during BBN. If DM has a thermal abundance and is relativistic at a SM temperature of a few MeV, it contributes to the number of relativistic degrees of freedom in the early universe modifying $N_{\rm{eff}}$ and changing the abundances of light elements like helium. Measurements of the helium fraction can thus be used to place a lower bound on the mass of thermal DM, $m_\chi \gtrsim 10\,\mathrm{MeV}$~\cite{Depta:2019lbe,Sabti:2019mhn}. In this work, we conservatively consider DM above this scale to ensure that the parameter space is not ruled out. However, we note that this constraint could be slightly weaker in parts of our parameter space due to the early kinetic decoupling of DM from the SM. In particular, if the DM decouples early enough, then some of the SM degrees of freedom in the bath at that time, given by $g_{*,0}$ in total, heat the SM bath at later times and raise the SM temperature relative to $T_\mathrm{DM}$ by a factor of a few in order to conserve entropy, $T_\text{SM}/T_\mathrm{DM} \sim (g_{*,0} / g_{*, \text{BBN}})^{1/3}$. This means that the DM contribution to the energy density, and hence $N_\text{eff}$, could be diluted by a factor of $(g_{*,0} / g_{*, \text{BBN}})^{4/3}$. Furthermore, once the DM becomes non-relativistic at a temperature $T_\mathrm{DM} \sim m_{\chi}$, the DM temperature will drop even further relative to the SM temperature, $T_\mathrm{DM} \sim T_\text{SM}^2 (g_{*,0} / g_{*, \text{BBN}})^{-2/3} /m_{\chi}$ for $T_\text{SM}<m_{\chi}$. Therefore, by the time of BBN, MeV-scale DM may have had its energy density diluted and may be non-relativistic, thus not contributing substantially to $N_\text{eff}$. In concert, these effects could increase the range of allowed masses for this model; we leave a more detailed exploration to future work.

In addition to the modification of $N_\mathrm{eff}$, DM annihilating into SM states at a temperature of a few keV can inject energy into the SM plasma causing the photo-disassociation of light  nuclei like deuterium. The corresponding bound on the annihilation cross-section not only depends on the DM mass and the relevant final states but also depends on the temperature of the kinetic decoupling, $T_\mathrm{kd}$, in the case when the thermally averaged cross-section has a temperature dependence \cite{Depta:2019lbe}. In our model, the cross-sections show this dependence around the keV-scale temperatures relevant for photo-disassociation for $\epsilon_R\ll 1$. Accurately evaluating this bound for such small $\epsilon_R$'s is therefore non-trivial, and we leave a detailed study for future work. For the purposes of this work, we note that for a velocity-independent annihilation cross-section, the bound from photo-disassociation corresponds to $\sigma v \lesssim \mathrm{few} \times 10^{-25} \mathrm{cm^3/s}$, and the constraint for velocity dependent cross-section gets weaker with increasing $T_\mathrm{kd}$ \cite{Depta:2019lbe}. For resonant annihilations, even the small couplings considered in this work can result in significantly larger cross sections at the late times relevant to BBN. Since the resonant annihilation cross section peaks near $T \sim \epsilon_R m_\chi $, in this work we consider $\epsilon_R \geq 0.001$, which along with the conservative lower bound on $m_\chi$ discussed above, ensures that the DM annihilation cross section is always below the upper bound during temperatures relevant for photo-disassociation. 

\subsection{Cosmic Microwave Background}
DM annihilation into SM final states during recombination can inject energy into the SM plasma. This energy injection can alter the ionization history and affect CMB anisotropies due to the scattering of CMB photons on additional free electrons that would have otherwise recombined into neutral atoms in the standard cosmology. This injection is usually described in terms of an effective parameter~\cite{Planck:2018vyg}, 
\begin{align}
\label{eq:CMBbound}
    p_\mathrm{ann} = 2f^\ast f_\mathrm{eff} \frac{\langle\sigma v \rangle_\mathrm{CMB} }{m_\chi}< 3.2 \times 10^{-28} \mathrm{cm}^3\mathrm{s}^{-1}\mathrm{GeV}^{-1}\,,
\end{align}
where $f^\ast$ is the fraction of DM in the excited state as described in the previous Section, $f_\mathrm{eff}$ is the efficiency fraction of injected energy that gets deposited in the plasma (where we adopt the values from Ref.~\cite{Bernreuther:2020koj} using the spectra from Refs.~\cite{Cirelli:2010xx,Coogan:2022cdd} and which depend on the DM mass and SM final state), and $\langle\sigma v \rangle_\mathrm{CMB}$ is the total annihilation cross section into SM states at recombination. The bound on $p_\text{ann}$ includes annihilation channels into all visible SM final states (i.e. excluding neutrino final states). In our case, the final states are dominantly electrons (and muons for DM masses above the muon mass), and we include the relevant branching fractions as appropriate when computing the total annihilation cross section. In the case of multiple possible final states, we weight the cross sections by the relevant deposited energy efficiency fraction, $f_\mathrm{eff}$. Since the DM particles are highly non-relativistic during and after recombination, to obtain $\langle \sigma v\rangle_\mathrm{CMB}$ we evaluate Eq.~\eqref{eq:sigmavthav} in the limit $\epsilon\to0$, 
\begin{align}
\left\langle \sigma v\right\rangle _{\mathrm{CMB}}=\frac{\alpha g_{\chi}^{2}\kappa^{2}}{s_0^2 }\frac{(s_0+2m_{e}^{2}) \sqrt{\left(s_0-4m_{e}^{2}\right)\left(s_0 - \delta^2\right)}}{B_{e}\left(\sqrt{s_0}\right)\left((1+\epsilon_{R})\Gamma_{A'}^{2}+s_0\epsilon_{R}^{2}\right)}.
\label{cmbann}
\end{align}

To compute the CMB limit on our parameter space, we require that the total $f_\mathrm{eff}$-weighted cross section in Eq.~\eqref{cmbann} not exceed the one implied by the bound on $p_\text{ann}$. The excluded parameters are depicted in Fig.~\ref{kappavsgx} as dotted lines which, despite being excluded by the CMB, would yield the observed amount of DM, as described in the previous Section. We find that -- in contrast to the case of non-resonant pseudo-Dirac DM -- very small mass splittings and large excited state fractions remain unconstrained by the CMB, particularly in the part of the plane corresponding to small couplings. Additionally, for the sub-keV values of $\delta$ that we consider here, the bound from the CMB is independent of $\delta$ because we do not get a substantial enough suppression in the excited state abundance as a result of the early kinetic decoupling (i.e. $f^\ast\sim 1$ near the boundary of the CMB exclusion). 

\subsection{Self-interacting DM}
Models of inelastic DM can also have unique SIDM behaviour in DM halos, affecting density profiles and subhalo mass functions in a way that is distinct from purely elastic SIDM~\cite{Vogelsberger:2018bok,ONeil:2022szc}. In this model, elastic scattering between the ground and excited states can occur at tree level, which is especially relevant in the thermal histories we consider where it is possible to have a high abundance of the excited state at late times. As we are in the Born regime, $\alpha_\chi m_\chi/m_{A'} \approx \alpha_\chi/2 \ll 1$, we can use results previously derived in the literature for the relevant SIDM cross sections. The $s$-channel resonance is not typically relevant in astrophysical environments (in contrast to e.g. Ref.~\cite{Chu:2018fzy}), since the lowest value of $\epsilon_R$ we consider is $\sim10^{-3}$ due to the strong BBN constraints described in Sec.~\ref{sec:bbn}, which corresponds to a minimum resonant velocity of $\sim 10^4$~km/s, in contrast to the $\sim 10^2$~km/s velocities that are typical in galaxies like the Milky Way. 

For tree-level elastic scattering between the ground and excited states, we employ the Born cross section of Ref.~\cite{Feng:2009hw}, assuming that the $t$-channel dominates. We find that the elastic scattering cross section is generally much less than the $\sigma/m_\chi \sim 1$~cm$^2$/g characteristic of SIDM constraints from merging galaxy clusters~\cite{Tulin:2017ara}. The exception to this lies at the edge of perturbativity $g_\chi \sim 1$ for the lightest DM masses we consider, $m_\chi \sim 10$~MeV. However, this part of the parameter space is excluded by the CMB, as evident from Fig.~\ref{kappavsgx}. Even if the CMB constraint could be circumvented, in this part of the parameter space the abundance of excited state particles (which would be required for tree-level elastic scattering) is generally more suppressed as shown in Fig.~\ref{fig:fstar}, which would also weaken the bounds from merging clusters. Elastic scattering between particles in the same mass state (i.e. two ground state particles) only occurs at the 1-loop level in this model; we confirm that the cross sections for these processes fall well below the $\sim 1$~cm$^2$/g level (due to the $\alpha_\chi^4$ scaling) using the expressions in the Supplemental Materials of Ref.~\cite{Fitzpatrick:2021cij}. We therefore conclude that merging cluster constraints on elastic SIDM do not have any impact in the viable parameter space for this model. 

Upscattering from the ground state to the excited state is kinematically forbidden below the velocity threshold of $v\sim  \sqrt{\delta/m_\chi}$, which can take on a wide range of values in the parameter space we consider, some of which are relevant to astrophysical systems. Above this velocity threshold, the cross section for upscattering in the Born regime saturates to the value of the elastic cross section between the ground and excited states~\cite{Schutz:2014nka}; while this cross section is generally below $\sim 1$~cm$^2$/g in our parameter space, the endothermic nature of the scattering can lead to substantial qualitative differences in the DM distribution compared to the elastic scattering case~\cite{ONeil:2022szc}, and therefore strong conclusions cannot be drawn about observational prospects without dedicated simulation work. Similarly, downscattering from the excited state to the ground state takes on the same value as upscattering once the velocities are above threshold; below threshold, there is an enhancement to the downscattering cross section, rendering it potentially quite large at low velocities (with $\sigma/m_\chi \gg 1$~cm$^2$/g). The effects of inelastic scattering has only begun to be explored in simulation, with no direct analog of this situation having been analyzed. In this thermal history, up to 50\% of the DM begins in the excited state similar to Ref.~\cite{Vogelsberger:2018bok}, which showed that even a few percent of the DM downscattering can have a significant effect on the structure of DM halos. On the other hand, the velocity thresholds in our parameter space can be easily accessible in astrophysical systems, potentially leading to some upscattering, which has a highly nontrivial interplay with the effects of downscattering as studied in Ref.~\cite{ONeil:2022szc}. Clearly futher exploration of the parameter space in simulation will be fruitful for connecting late-universe halo observables to DM parameters motivated by self-consistent DM thermal histories. 

\subsection{Indirect Detection}
\begin{figure}
\includegraphics[width=0.49\textwidth]{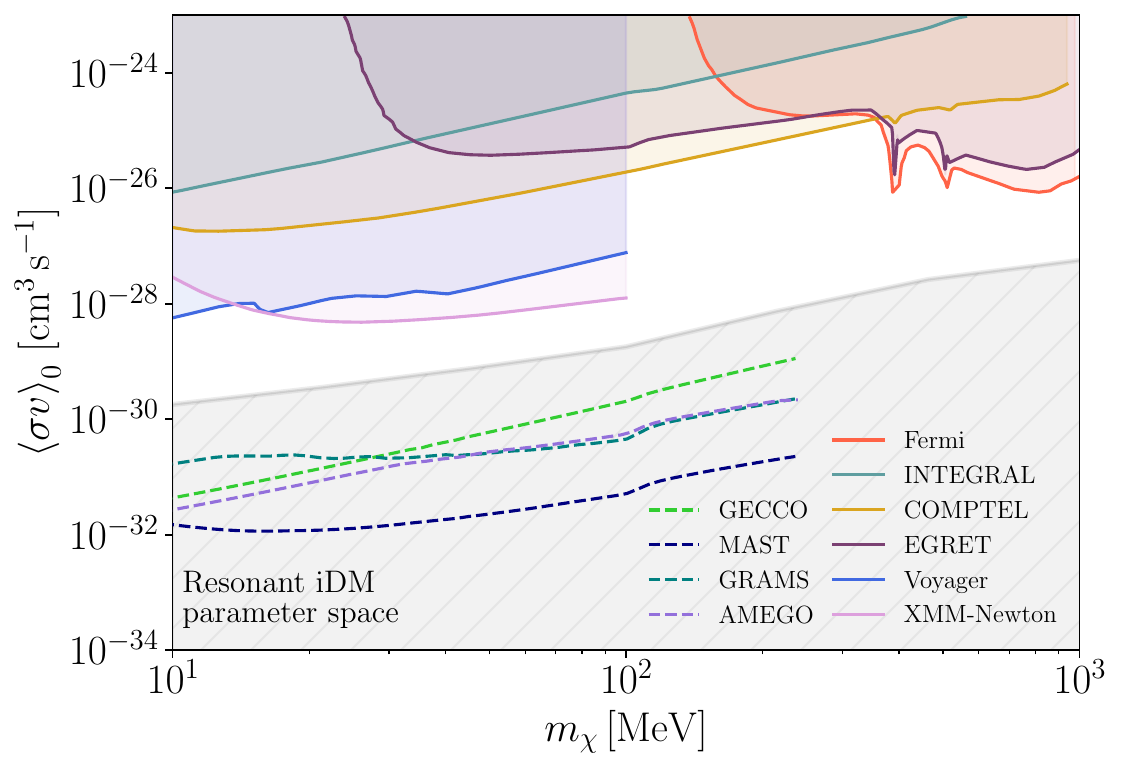}
\caption{Bounds on the total DM annihilation cross-section as a function of DM mass from various $\gamma-$rays and X-rays experiments~\cite{Coogan:2022cdd, Boudaud:2016mos, Cirelli:2023tnx} (solid lines) as well as projections \cite{Orlando:2021get,Coogan:2021rez,Dzhatdoev:2019kay,Aramaki:2019bpi,Aramaki:2020gqm,AMEGO:2019gny,Kierans:2020otl} (dashed lines). The resonant iDM parameter space allowed by the CMB is shown in grey.\label{fig:indirect}}
\end{figure}
\begin{figure*}
\centering
\includegraphics[width=\textwidth]{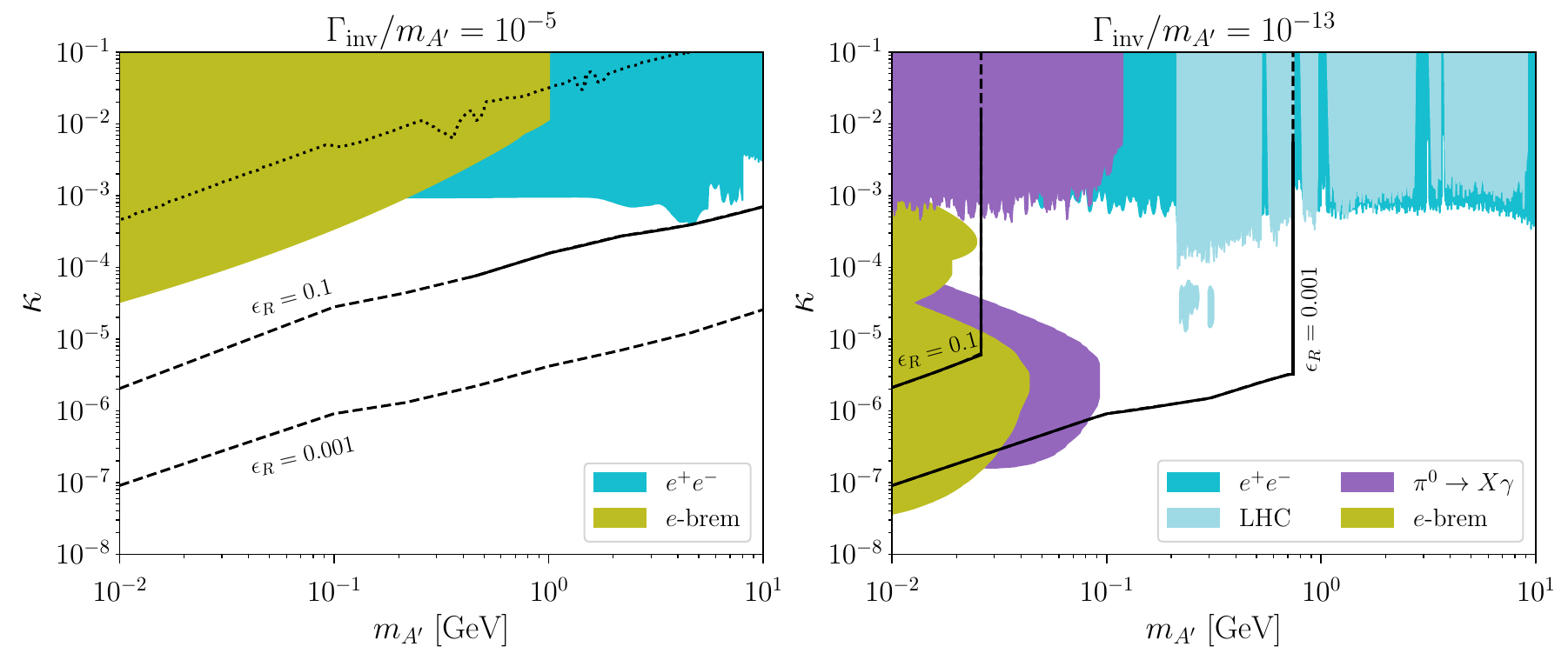}
\caption{\label{fig:acc} Some of the strongest constraints in the $\kappa-m_{A^\prime}$ plane from  accelerator experiments, including BaBar~\cite{Lees:2017lec,Lees:2014xha,TheBABAR:2016rlg}, NA64~\cite{NA64:2019imj, Banerjee:2019hmi}, LHCb~\cite{Aaij:2017rft, Aaij:2019bvg}, CMS~\cite{CMS:2019kiy}, NuCal~\cite{Tsai:2019mtm, Blumlein:1990ay}, E141~\cite{Riordan:1987aw}, NA48~\cite{Batley:2015lha}, and E137~\cite{Andreas:2012mt}, as computed with \texttt{DarkCast}~\cite{Ilten:2018crw}. The different colours correspond to different dark photon production channels.
We show two representative cases where the dark photon decays primarily invisibly (left) and visibly (right). The solid black lines represent the target for $\epsilon_R = 0.001$ and $\epsilon_R=0.1$, with the dashed portions corresponding to exclusions from the CMB. Also shown for comparison is the thermal prediction assuming thermal freeze-out with $m_{A^\prime} = 3 m_\chi$ (dotted line)~\cite{Berlin:2018bsc}.}
\end{figure*}
DM annihilation in astrophysical environments produces cosmic rays and high-energy photons, which can be employed to search indirectly for DM. A null detection of annihilation byproducts can thus be used to constrain the annihilation cross section, $\langle \sigma v \rangle $. The strength of the expected signal additionally depends on the integrated DM density (the $J$ factor). As discussed in the previous Subsection, SIDM effects, particularly those from inelastic interactions, may alter the expected density profiles of DM halos. As this has yet to be quantified via simulation, we assume in this discussion that resonant inelastic DM has the same $J$ factor as what is typically considered in the literature. Additionally, as shown in Sec. \ref{sec:cosmo_const}, the CMB constraints on DM annihilation push us toward $g_\chi \lesssim 0.1$, which corresponds to the ground and excited states being equally populated, or $f^\ast \sim 1$, for the $\delta$ ranges of interest (see also Fig.~\ref{fig:fstar}). Therefore, as opposed to standard thermal histories of inelastic DM, we do not necessarily have a late-time suppression of annihilation from the thermal depletion of the excited state. 

Since the DM velocity in present-day halos is too small (by about two orders of magnitude) for the dark photon mediator to be produced on shell, the DM annihilation rate at late times can be calculated in the heavy mediator (non-resonant) limit. In Fig. \ref{fig:indirect}, we show the constraints on the \textit{total} present-day DM annihilation cross-section to SM states, $\langle \sigma v \rangle_0$, from measurements of the gamma-ray flux from the galactic center using Fermi INTEGRAL, COMPTEL,  and EGRET~\cite{Coogan:2022cdd}. We also show constraints from Voyager~\cite{Boudaud:2016mos} and XMM-Newton~\cite{Cirelli:2023tnx}, which bound the DM annihilation cross-section to electrons and therefore have been cut off at the muon threshold, $m_\mu \sim 100 \,\mathrm{MeV}$ beyond which annihilation to other final states becomes relevant. For reference, we show the resonant pseudo-Dirac DM parameter region allowed by the CMB in grey. Even though current constraints do not yet probe the target parameter space that is allowed by the CMB, future telescopes such as GECCO~\cite{Orlando:2021get,Coogan:2021rez}, MAST~\cite{Dzhatdoev:2019kay}, GRAMS~\cite{Aramaki:2019bpi,Aramaki:2020gqm} and AMEGO~\cite{AMEGO:2019gny,Kierans:2020otl} will be able to explore the parameter space~\cite{Coogan:2021sjs}. 

\section{Terrestrial searches}
\label{sec:experiment}
\subsection{Accelerator searches}
Accelerator experiments provide a complementary probe to search for DM and any associated mediators. Dark photons with masses in the MeV-GeV range can be produced at collider and beam dump experiments, resulting in either missing transverse energy or displaced vertex signatures. The production cross-section for the dark photons depends on their coupling to the SM, $\kappa$, whereas the decay signature depends on their lifetime, $\tau_{A^\prime}=1/\Gamma_{A^\prime}$ which is in general a function of $\kappa, \,g_\chi,\,\epsilon_R$ and $\delta$. Since $\delta \ll m_{A^\prime},\, m_{\chi}$, one can assume to a very good approximation that $\Gamma_{A^\prime}$ is independent of $\delta$ (i.e. we can take $\delta \to 0$). In this limit, the bounds presented in Ref.~\cite{Bernreuther:2020koj} which were obtained using a modified version of \texttt{Darkcast} \cite{Ilten:2018crw} can be directly applied to our parameter space. In particular, as was pointed out in Ref.~\cite{Bernreuther:2020koj}, for a given $m_{A^\prime}$, the constraints on $\kappa$ depend only on a combination of $g_\chi$ and $\epsilon_R$ through the dark photon's reduced invisible decay width, 
\begin{align}
    \gamma_\mathrm{inv} \equiv \frac{\Gamma_{A^\prime}}{m_{A^\prime}} = \frac{g_\chi^2}{12}\left(1-\frac{1}{1+\epsilon_R}\right)^{1/2}\left(1+\frac{1}{2(1+\epsilon_R)}\right)\,.
\end{align}

\begin{figure*}[t!]
\centering
\includegraphics[width=\textwidth]{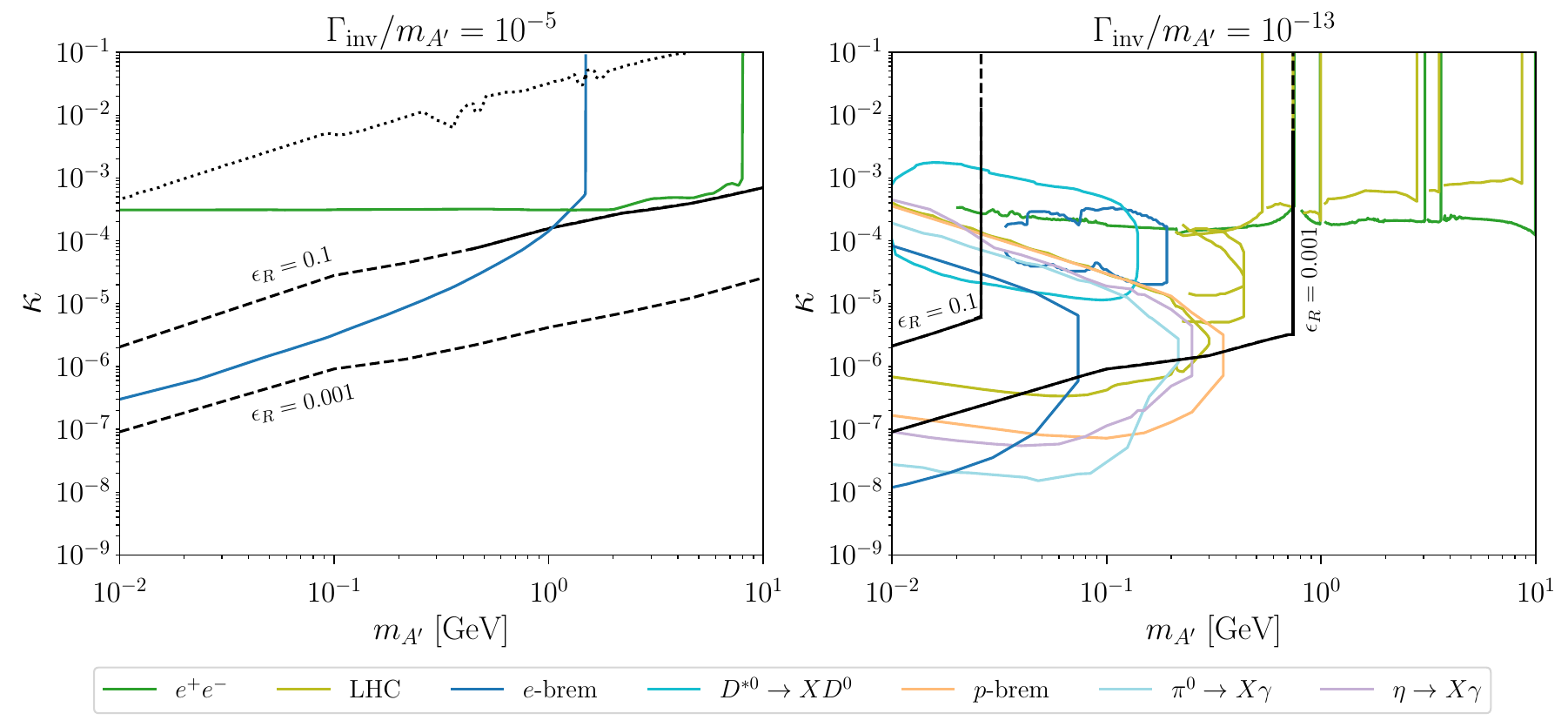}
\caption{\label{fig:acc_future}
Same as Fig.~\ref{fig:acc} except with some of the strongest projections for future experiments including Belle~II~\cite{Kou:2018nap}, FASER~\cite{Ariga:2018uku}, HPS~\cite{Baltzell:2016eee}, LDMX~\cite{LDMX:2018cma}, LHCb~\cite{Ilten:2015hya,Ilten:2016tkc}, SeaQuest~\cite{Gardner:2015wea}, SHiP~\cite{Alekhin:2015byh, Ahdida:2020new}, and Yemilab~\cite{Seo:2020dtx}.}
\end{figure*}

In Fig. \ref{fig:acc}, we display the bounds on dark photons in the $\kappa-m_{A^\prime}$ plane for two fixed values of the dark photon's reduced invisible width, $\gamma_\mathrm{inv}= 10^{-5}$ (left) and $\gamma_\mathrm{inv}= 10^{-13}$ (right). The shaded regions correspond to different dark photon production channels. For $\gamma_\mathrm{inv}= 10^{-5}$, the invisible decay width is larger than the visible one for $\kappa \lesssim 0.1 $ and therefore the beam dump experiments that search for $A^\prime$ decays to leptons lose sensitivity. The solid black lines in both panels correspond to the $\kappa$ values that reproduce the observed DM abundance following the production mechanism outlined in Sec. \ref{sec:cosmo}, for $\epsilon_R\in[0.001,\,0.1]$. The dashed parts of the black lines correspond to the points excluded by the CMB. We note that for $\Gamma_\mathrm{DM}>\Gamma_\mathrm{SM}$, (or equivalently $g_\chi \gg \kappa$), the CMB constrains the part of the parameter space which cannot be probed by accelerator experiments making the accelerator and cosmological bounds highly complementary. Additionally, we find that for a given $\gamma_\mathrm{inv}$ and $\epsilon_R$, or equivalently, a fixed $g_\chi$, there exists a maximum $m_{A^\prime}$ beyond which DM is always over-produced. From Eq.~\eqref{eq:oh2scale}, we see that the DM relic density is proportional to $m_{A^\prime}/\min(g_\chi^2,\,\kappa^2)$ meaning that once the couplings are fixed, a larger $m_{A^\prime}$ results in a larger DM abundance. This causes the relic lines to be vertical in the right panel of Fig. \ref{fig:acc} (in the left panel, the maximum $m_{A^\prime}$ lies outside the plotted region). Finally, we note that for a given $\gamma_\mathrm{inv}$, $\kappa$-values bounded by the two curves corresponding to $\epsilon_R=0.001$ and $\epsilon_R=0.1$ also reproduce the observed DM abundance for different values of $\epsilon_R$ resulting in a much broader thermal target. For reference, we show the usual thermal target assuming non-resonant, $s$-wave thermal freeze-out (dotted black line) in the left panel of Fig.~\ref{fig:acc}. Future accelerator experiments are poised to more fully explore the parameter space of this model, as shown in Fig.~\ref{fig:acc_future}

\subsection{Direct Detection}
Most thermal histories for pseudo-Dirac DM result in a relic excited state fraction suppressed by several orders of magnitude such that most of the DM in the halo of the Milky Way (MW) is in the ground state. In this case, because of the off-diagonal coupling, the only tree-level scattering process on a SM target would be upscattering to the excited state. In contrast, for the thermal history considered in this work, around half of the DM is in the excited state at late times for most parts of the parameter space. In particular, the largest values of $g_\chi$ that suppress the abundance of $\chi_2$ in this thermal history are already excluded by the CMB (see Figs. \ref{kappavsgx} and \ref{fig:fstar}). Therefore, the primary signature of pseudo-Dirac DM in direct detection experiments is downscattering of the excited state $\chi_2 e \to \chi_1 e$, which deposits an energy $\sim \delta$. The deposited energy can ionize electrons in the target which can be detected either directly through charge-coupled devices~\cite{SENSEI:2021hcn}, or by detecting secondary scintillation photons using photomultiplier tubes \cite{XENON:2019gfn}. The absence of a kinematic barrier for downscattering implies an enhancement in the event rate for sub-GeV DM.

In the following discussion, we consider DM-electron scattering in semiconductor and Xenon targets, and we place constraints on the fiducial DM-electron scattering cross-section \cite{Essig:2011nj},
\begin{align}
    \overline{\sigma}_e = \frac{4\mu_{\chi,e}^2\alpha \kappa^2 g_\chi^2}{(m_{A^\prime}^2 + \alpha^2 m_e^2)^2}\,,
\end{align}
where $\mu_{\chi,e}$ is the reduced mass of the DM-electron system. The recoil energy of the electron is $E_{Re}=\Delta E_{e} - \Delta E_{B}$ where $\Delta E_{B}$ is the electron binding energy and the energy deposited by sub-GeV DM downscattering is \begin{equation}
    \Delta E_{e}=\mathbf{q}\cdot \mathbf{v}-\frac{\mathbf{q}^{2}}{2 m_{\chi }}+\delta
\end{equation}
for momentum transfer $\textbf{q}$ and relative velocity $\textbf{v}$. For downscattering, the minimum velocity required to transfer a momentum with magnitude $q$ and an energy $\Delta E_e$ to the electron is therefore given by 
\begin{align}
    v_{\rm{min}}\left(q,\Delta E_e \right) \equiv
\left|\frac{\Delta E_{e}-\delta}{q}+\frac{q}{2m_{\chi}}\right|,
\end{align}
which corresponds to the differential event rate for atomic targets~\cite{Essig:2011nj, Bloch:2020uzh},
\begin{align}
\label{eq:rateeq}
    \frac{\mathrm{d}R}{\mathrm{d}\Delta E_e}& = \frac{\overline{\sigma}_e}{8\mu_{\chi e}^2}\sum_{n,l}(\Delta E_e - E_{nl})^{-1}\frac{\rho_{\chi_2}}{m_{\chi_2}}  \\
    &\times \int q\mathrm{d}q |f_{nl \to \Delta E_e - E_{nl}}(q)|^2\eta(v_{\rm{min}}(q, \Delta E_e))\,, \nonumber
\end{align}
where $\Delta E_{n,l}$ is the binding energy of the electron in the $nl$ shell, $\rho_{\chi_2}=f^\ast \rho_{\mathrm{DM}} \approx 0.5\times 0.4 ~\mathrm{GeV}~\mathrm{cm}^{-3}$ is the density of the excited state~\cite{ParticleDataGroup:2018ovx}, and $f_{nl \to \Delta E_e - E_{nl}}(q)$ is the electron ionization form factor which we evaluated numerically following the prescription of Ref.~\cite{Catena:2019gfa} using \texttt{DarkART} \cite{DarkART}. The function $\eta(v_\text{min})$ in Eq.~\eqref{eq:rateeq} can be related to the DM-velocity distribution $f_{\chi}(v)$ by 
 \begin{equation}
\eta(v_{min})\equiv\int\frac{d^{3}v}{v}f_{\chi}(v)\Theta\left(v-v_{min}\right).
 \end{equation} 
In this work, we assume the Standard Halo Model for $f_\chi (v)$ \cite{Evans:2018bqy}. The magnitude of the momentum transferred to the electron for a given DM mass and energy deposition $\Delta E_e$ is bounded by 
\begin{align}
    q_\text{min} &=\text{sign}(\Delta E_{e}-\delta)m_{\chi}v_\text{max}\left(1-\sqrt{1-\frac{\Delta E_{e}-\delta}{\frac{1}{2}m_{\chi}v_\text{max}^{2}}}\right) \nonumber\\
q_\text{max} &=m_{\chi}v_\text{max}\left(1+\sqrt{1-\frac{\Delta E_{e}-\delta}{\frac{1}{2}m_{\chi}v_\text{max}^{2}}}\right)
\end{align}
where $v_{\rm{max}}$ is the largest DM velocity relative to the detector frame as determined by the escape velocity of the halo at the Earth's position and the velocity of Earth in the Galactic rest frame \cite{Evans:2018bqy}. 
\begin{figure}
\centering
\includegraphics[width=0.48\textwidth]{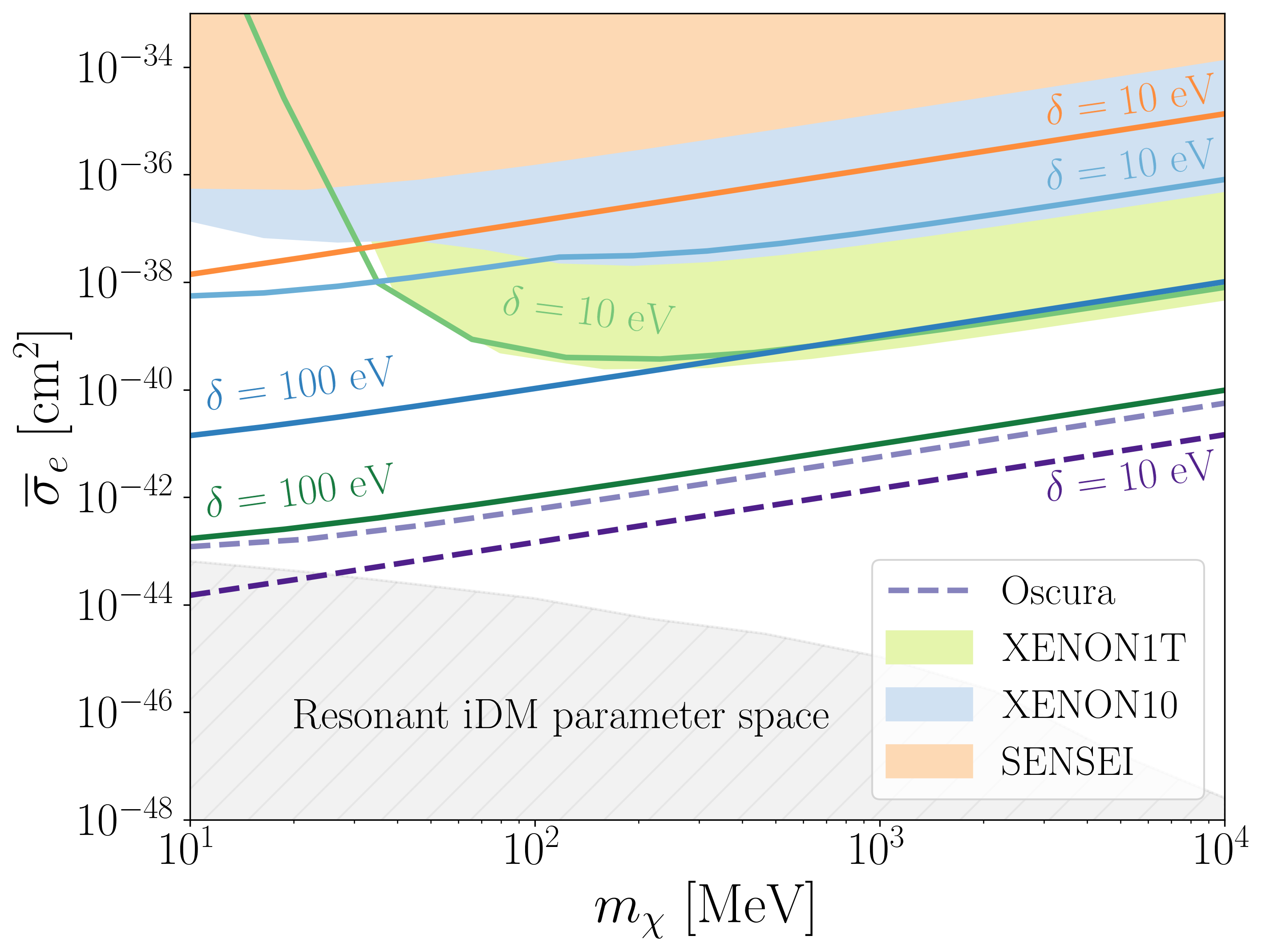}
\caption{\label{fig:DD} Constraints on the DM-electron scattering cross-section, $\sigma_e$ as a function of the DM mass from XENON10 (blue), XENON1T (green), and SENSEI (orange) for different values of $\delta$, with $f^\ast=1$. The shaded regions correspond to exclusions on elastic scattering, i.e., $\delta = 0$. The dashed purple lines are sensitivity curves for Oscura assuming $\delta=0$ (light) and $\delta=10\,\rm{eV}$ (dark) respectively. In grey, we show the resonant iDM parameter space allowed by CMB.}
\end{figure}
A similar expression as Eq.~\eqref{eq:rateeq} can be obtained for semiconductor targets which depends instead on the crystal form factor as described in Ref.~\cite{Essig:2015cda}. The relevant energy threshold for such a target is the band gap between the valence band to the conduction band. Note that in Eq.~\eqref{eq:rateeq}, we have set the DM form factor to unity, $F_{\rm{DM}}=1$, which corresponds to the heavy mediator limit, $m_{A^\prime} \gg \alpha m_e$ as determined by the typical Fermi momentum (for semiconductors targets) or inverse Bohr radius (for atomic targets). 

In contrast to the case of elastic scattering, the minimum momentum transfer, $q_{\rm{min}}$ for a given energy transfer $\Delta E_e$, can be zero if the mass splitting is above the energy threshold for ionization. This results in much larger event rates since the inelastic scattering kinematics have substantial overlap with peaks in the electron ionisation form factor \cite{Emken:2021vmf}. A similar effect also occurs for semiconductor targets where the peaks in the crystal form factor become more kinematically accessible for $\delta>0$~\cite{Essig:2015cda}.
Inelastic scattering also results in a characteristic spectrum of events peaked around $\Delta E_e \sim \delta$. As a result, bounds on $\overline{\sigma}_e$ derived under the assumption of elastic scattering cannot be directly applied to the parameter space of this model.

In order to re-cast the bounds, we use the prescription outlined in Ref.~\cite{Essig:2017kqs} for calculating event rates in the XENON10 ~\cite{XENON10:2011prx, Essig:2017kqs}, XENON1T ~\cite{XENON:2019gfn} and SENSEI ~\cite{SENSEI:2020dpa} experiments. For XENON10 and XENON1T, we calculate the electron ionization form factors using \texttt{DarkART}~\cite{DarkART}. The crystal form factors for SENSEI are obtained from \texttt{QEDark}~\cite{Essig:2015cda}. We use the publicly available data for the three experiments~\cite{XENON10:2011prx, Essig:2017kqs,XENON:2019gfn,SENSEI:2020dpa} to derive 90\% confidence level exclusions on $\overline{\sigma}_e$, shown in Fig. \ref{fig:DD}. For XENON10 and XENON1T, we show the exclusion for $\delta = 10\, \rm{eV}$ and $\delta = 100\,\rm{eV}$, while for SENSEI, we show the exclusion for $\delta = 10 \,\rm{eV}$. We verify that our analysis reproduces the elastic scattering ($\delta = 0$) bounds from these experiments, shown as shaded regions in Fig. \ref{fig:DD}. We note that these bounds are conservative since we do not model any backgrounds or place any cuts in the observed events. In other words, we treat any event as a potential DM signal, resulting in the weakest possible limits on the cross section. 

As shown in Fig. \ref{fig:DD}, the allowed parameter space of this model (represented by the grey band) is an attractive target for upcoming experiments probing light DM. In particular, we plot the sensitivity curves for Oscura \cite{Oscura:2022vmi} for $\delta = 0\,\rm{eV}$ (light purple) and $\delta = 10\,\rm{eV}$ (purple) assuming an exposure of 30 kg-year. Furthermore, for the $\mathcal{O}$(eV) values of $\delta$ we consider here, up-scattering inside the Earth may result in an enhanced density of the excited state at the detector, which would also result in stronger bounds and forecasts than the ones presented here~\cite{Emken:2021vmf}. Finally, even stronger bounds and forecasts may be obtained by considering electron ionization caused by DM-nucleon scattering through the Migdal effect \cite{Ibe:2017yqa,XENON:2019zpr, migdal}. These bounds were evaluated for $\delta \gtrsim O(\rm{keV})$ in Ref.~\cite{Bell:2021zkr}. We leave an analysis of these bounds for sub-keV values of $\delta$ for future work.

\section{Discussion}
\label{sec:conclusion}
Pseudo-Dirac DM is a minimal modification of standard vector portal DM that can result in qualitatively new cosmological, astrophysical, and experimental phenomenology. In this work, we examine the parameter space of this well-studied model in the regime of small (sub-keV) mass splittings and in the presence of resonant annihilations, $\chi_2\chi_1\to A^\prime \to \mathrm{SM}~\mathrm{SM}$, where $m_{A^\prime} \approx 2 m_\chi$. The resonantly enhanced annihilations imply that tiny couplings are able to reproduce the observed DM relic density.
Additionally, because of these small couplings, DM can kinetically decouple from the SM before its final relic abundance is reached. Therefore, in order to accurately predict the relic density, one must solve a coupled system of Boltzmann equations for the densities and temperatures of the relevant species. We used the numerical Boltzmann solver \texttt{DRAKE} to properly account for this effect and found that the predicted DM abundance can have corrections as large as an order of magnitude, depending on the underlying parameters of the theory.

The early kinetic decoupling ensures that the excited state is not thermally depleted. Despite the presence of the excited state, this model is consistent with strong bounds coming from BBN and the CMB owing to the strong velocity suppression in the annihilation cross-section at sub-keV temperatures (when the dark photon can no longer be produced on-shell). As a result, as shown in Fig. \ref{kappavsgx}, most of the parameter space of this model is unconstrained, in contrast to sub-GeV Dirac DM that freezes-out through an $s$-wave process. 

The presence of the long-lived excited state can have unique astrophysical signatures that are usually not relevant for pseudo-Dirac DM. For instance, tree-level elastic scattering could cause SIDM behavior, with the caveat that cross sections exceeding $\sigma/m_\chi\sim 1$~cm$^2$/g either have couplings that are excluded by the CMB or lie in a region of parameter space with $f^\ast \sim 0.01$. More notably, exothermic downscattering can be relevant, especially in low-velocity environments where there is an enhancement to the cross section. The extent to which exothermic scattering matters \emph{in situ} is difficult to quantify without further simulation of inelastic SIDM halos, however previous work has found that small mass splittings with $\delta/m_\chi \sim 10^{-6}$ can have a dramatic impact on the properties of a DM halo and its subhalos~\cite{ONeil:2022szc, Vogelsberger:2018bok}. 
The relic excited state can also result in signals in indirect detection experiments. The ground and excited state present in the MW can annihilate into various SM states and therefore gamma-ray and X-ray telescopes can be used to look for signatures of this model. Since DM annihilation at late times mimics the off-resonance Dirac case, we use previous analyses of gamma-ray and $X$-ray data \cite{Coogan:2022cdd,Cirelli:2023tnx,Boudaud:2016mos} to place bounds on the total DM annihilation cross-section. We find that despite the small couplings, the resonant inelastic parameter space is an attractive target for the next generation of telescopes such as GECCO, MAST, GRAMS and AMEGO (see Fig. \ref{fig:indirect}).

The excited state can downscatter in a range of direct detection experiments. Because of the absence of kinematic barrier for this process, the event rate is significantly enhanced compared to elastic scattering and also compared to pseudo-Dirac DM thermal histories with an exponentially suppressed abundance of the excited state. Using state-of-the-art numerical codes \texttt{DarkART} and \texttt{QEDark} to obtain the electron ionisation and crystal form factors respectively, we calculate the event rates for inelastic DM-electron scattering at Xenon- and Silicon-based experiments. We use the analysis procedure described in Ref.~\cite{Essig:2017kqs} to place bounds on the DM-electron scattering cross-section for different values of $\delta$. We find that future semiconductor-based experiments such as Oscura will begin to probe the resonant inelastic DM parameter space as shown in Fig. \ref{fig:DD}.

Simultaneously, accelerator searches for dark photons can also explore relevant parameter space for this model. We use the publicly available code \texttt{Darkcast}, to depict bounds on the kinetic mixing, $\kappa$ for fixed values of $g_\chi$. The presence of a resonance implies a broadening of the thermal target as shown in Fig.~\ref{fig:acc}. We find that accelerator bounds are complementary to those set by the CMB, as the accelerator experiments probe the parts of the parameter space that are harder to constrain using early-universe probes. Future experiments will constrain large parts of the parameter space of this model, as shown in Fig.~\ref{fig:acc_future}. 

In summary, sub-GeV resonant pseudo-Dirac DM is an attractive thermal target for a variety of terrestrial DM experiments and astrophysical searches. The complete exploration of the phenomenology of this model leaves a lot of promising directions for future work. In particular, a more accurate treatment of the photo-disassociation bounds coming from BBN may further constrain this parameter space. Additionally, the early kinetic decoupling present in this model may loosen the $m_\chi \geq 10\,\mathrm{MeV}$ lower bound on the mass of thermal DM coming from $N_\mathrm{eff}$, despite the thermal equilibrium between the dark and visible sectors at early times. Furthermore, it will be necessary to perform additional cosmological simulations  in order to understand the effect of up- and downscattering on halo properties which have immediate consequences for direct and indirect detection experiments. Finally, a more rigorous analysis of the direct detection bounds needs to be undertaken, including (1) the upscattering of the ground state as it passes through the Earth before downscattering in the detector and (2) scattering on electrons through the Migdal effect. Such an analysis may result in even stronger bounds and forecasts on resonant pseudo-Dirac DM than the ones presented here. 

\section*{Acknowledgements}
It is a pleasure to thank Daniel Baxter, Asher Berlin, Elias Bernreuther, Timon Emken, Felix Kahlhoefer, Tongyan Lin, and Tien-Tien Yu for useful conversations and correspondence pertaining to this work. We especially thank Neal Weiner for useful comments on the manuscript. The research of NB was undertaken thanks in part to funding from the Canada First Research Excellence Fund through the Arthur B. McDonald Canadian Astroparticle Physics Research Institute. SH was supported in part by a Trottier Space Institute Fellowship. NB, SH, and KS acknowledge support from a Natural Sciences and Engineering Research Council of Canada Subatomic Physics Discovery Grant and from the Canada Research Chairs program.

\bibliography{main}
\end{document}